\documentclass[11pt]{article}
\usepackage{amssymb,amsmath,amsfonts}
\usepackage{graphicx}
\usepackage{graphics}
\usepackage{eepic,epsfig}

\begin{document}

\title{Casimir effect in de Sitter spacetime with compactified dimension}
\author{ A. A. Saharian$^{1,2}$\thanks{%
E-mail: saharian@ictp.it } and M. R. Setare$^{3}$\thanks{%
E-mail: rezakord@ipm.ir} \\
\textit{$^1$ Department of Physics, Yerevan State University, Yerevan,
Armenia } \\
\textit{$^2$ International Centre for Theoretical Physics, Trieste, Italy}\\
\textit{$^{3}$ Department of Science, Payame Noor University, Bijar, Iran}}
\maketitle

\begin{abstract}
We investigate the Hadamard function, the vacuum expectation values of the
field square and the energy-momentum tensor of a scalar field with general
curvature coupling parameter in de Sitter spacetime compactified along one
of spatial dimensions. By using the Abel-Plana summation formula, we have
explicitly extracted from the vacuum expectation values the part due to the
compactness of the spatial dimension. The topological part in the vacuum
energy-momentum tensor violates the local de Sitter symmetry and dominates
in the early stages of the cosmological evolution. At late times the
corresponding vacuum stresses are isotropic and the topological part
corresponds to an effective gravitational source with barotropic equation of
state.
\end{abstract}

\bigskip

\section{Introduction}

De Sitter (dS) spacetime is the maximally symmetric solution of Einsten's
equation with a positive cosmological constant. Recent astronomical
observations of supernovae and cosmic microwave background \cite{Ries07}
indicate that the universe is accelerating and can be well approximated by a
world with a positive cosmological constant. If the universe would
accelerate indefinitely, the standard cosmology leads to an asymptotic dS
universe. De Sitter spacetime plays an important role in the inflationary
scenario, where an exponentially expanding approximately dS spacetime is
employed to solve a number of problems in standard cosmology \cite{Lind90}.
In this regard in recent years, many string theorists have devoted to
understand and shed light on the cosmological constant or dark energy within
the string framework. The famous Kachru-Kallosh-Linde-Trivedi (KKLT) model
\cite{Kach03} is a typical example, which tries to construct metastable de
Sitter vacua in the light of type IIB string theory. The quantum field
theory on dS spacetime is also of considerable interest. In particular, the
inhomogeneities generated by fluctuations of a quantum field during
inflation provide an attractive mechanism for the structure formation in the
universe. Another motivation for investigations of dS based quantum theories
is related to the recently proposed holographic duality between quantum
gravity on dS spacetime and a quantum field theory living on boundary
identified with the timelike infinity of dS spacetime \cite{Stro01}.

In quantum field theory on curved backgrounds among the important quantities
describing the local properties of a quantum field and quantum back-reaction
effects are the expectation values of the field square and the
energy-momentum tensor for a given quantum state. In particular, the vacuum
expectation values of these quantities are of special interest. For a scalar
field on background of dS spacetime the renormalized vacuum expectation
values of the field square and the energy-momentum tensor are investigated
in Refs. \cite{Cand75,Dowk76,Dowk76b,Bunc78} by using various regularization
schemes. The corresponding effects upon phase transitions in an expanding
universe are discussed in \cite{Vile82}. Recently it was argued that there
is no reason to believe that the version of dS spacetime which may emerge
from string theory, will necessarily be the most familiar version with
symmetry group $O(1,4)$ and there are many different topological spaces
which can accept the dS metric locally (see \cite{Kach03,McIn04} and
references therein). There are many reasons to expect that in string theory
the most natural topology for the universe is that of a flat compact
three-manifold \cite{McIn04}. From an inflationary point of view universes
with compact spatial dimensions, under certain conditions, should be
considered a rule rather than an exception \cite{Lind04} (for observational
bounds on the size of compactified dimensions see Refs. \cite{Oliv96}).

The compactification of spatial dimensions leads to the modification of the
spectrum of vacuum fluctuations and, as a result, to Casimir-type
contributions to the vacuum expectation values of physical observables
(topological Casimir effect, see \cite{Most97}). In the present paper we
investigate the effect of the compactification of one of spatial dimensions
in dS spacetime on the properties of quantum vacuum for a scalar field with
general curvature coupling parameter (for quantum effects in braneworld
models with dS spaces see, for instance, \cite{dSbrane}).

The paper is organized as follows. In the next section we consider the
Hadamard function. By using the Abel-Plana summation formula, we decompose
this function in two parts: the first one is the corresponding function for
the uncompactified dS spacetime and the second one is induced by the
compactness of the spatial dimension. In Section \ref{sec:vevEmt} we use the
Hadamard function for the evaluation of the vacuum expectation values of the
field square and the energy-momentum tensor. As the parts corresponding to
the uncompactified dS spacetime are well-investigated in literature, we are
mainly concerned with the topological part. The asymptotic behavior of the
latter is investigated in detail in early and late stages of the
cosmological evolution. The main results of the paper are summarized in
Section \ref{sec:Conc}.

\section{Hadamard function in de Sitter spacetime with a compact spatial
dimension}

\label{sec:WF}

We consider a quantum scalar field with curvature coupling parameter $\xi $\
on background of the $(D+1)$-dimensional de Sitter spacetime. We will write
the corresponding line element in the form most appropriate for cosmological
applications:%
\begin{equation}
ds^{2}=dt^{2}-e^{2t/\alpha }\sum_{i=1}^{D}(dz^{i})^{2}.  \label{ds2deSit}
\end{equation}%
The parameter $\alpha $ is related to the Ricci scalar and the corresponding
cosmological constant by the formulae%
\begin{equation}
R=D(D+1)/\alpha ^{2},\;\Lambda =D(D-1)/2\alpha ^{2}.  \label{Ricci}
\end{equation}%
For the further discussion, in addition to the synchronous time coordinate $t
$ it is convenient to introduce the conformal time in accordance with%
\begin{equation}
\eta =\alpha e^{-t/\alpha },\;0<\eta <\infty .  \label{eta}
\end{equation}%
In terms of this coordinate the line element takes conformally flat form:%
\begin{equation}
ds^{2}=\alpha ^{2}\eta ^{-2}[d\eta ^{2}-\sum_{i=1}^{D}(dz^{i})^{2}].
\label{ds2Dd}
\end{equation}%
We will assume that the spatial coordinate $z^{D}$ is compactified to $S^{1}$%
: $0\leqslant z^{D}\leqslant L$. This geometry is a limiting case of
toroidally compactified dS spacetime discussed in \cite{Lind04} and can be
used to describe two types of models. For the first one $D=4$\ and it
corresponds to the universe with Kaluza-Klein type single extra dimension.
As it will be shown the presence of extra dimension generates an additional
gravitational source in the cosmological equations which is of barotropic
type at late stages of the cosmological evolution. For the second model $D=3$
and the results given below describe how the properties of the universe with
dS geometry are changed by one-loop quantum effects induced by the
compactness of a single spatial dimension.

The field equation has the form%
\begin{equation}
\left( \nabla _{l}\nabla ^{l}+m^{2}+\xi R\right) \varphi =0,  \label{fieldeq}
\end{equation}%
where $\nabla _{l}$ is the covariant derivative operator associated with
line element (\ref{ds2Dd}). The values of the curvature coupling parameter $%
\xi =0$ and $\xi =\xi _{D}\equiv (D-1)/4D$ correspond to the most important
special cases of minimally and conformally coupled fields. For a scalar
field with periodic boundary condition one has $\varphi (\eta ,\mathbf{z}%
,z^{D}+L)=\varphi (\eta ,\mathbf{z},z^{D})$, where $\mathbf{z}=(z^{1},\ldots
,z^{D-1})$ is the set of uncompactified dimensions. In this paper we are
interested in the effects of non-trivial topology on the VEVs of the field
square and the energy-momentum tensor. This VEVs are obtained from the
corresponding Hadamard function in the coincidence limit of the arguments.

To evaluate the Hadamard function we employ the mode-sum formula
\begin{equation}
G^{(1)}(x,x^{\prime })=\langle 0|\varphi (x)\varphi (x^{\prime })+\varphi
(x^{\prime })\varphi (x)|0\rangle =\sum_{\sigma }\left[ \varphi _{\sigma
}(x)\varphi _{\sigma }^{\ast }(x^{\prime })+\varphi _{\sigma }(x^{\prime
})\varphi _{\sigma }^{\ast }(x)\right] ,  \label{Hadam1}
\end{equation}%
where $\left\{ \varphi _{\sigma }(x),\varphi _{\sigma }^{\ast }(x)\right\} $
is a complete set of positive and negative frequency solutions to the
classical field equations and satisfying the periodicity condition along the
$z^{D}$- direction. The collective index $\sigma $ specifies these
solutions. For the problem under consideration the eigenfunctions have the
form
\begin{equation}
\varphi _{\sigma }(x)=C_{\sigma }\eta ^{D/2}H_{\nu }^{(2)}(k_{n}\eta )e^{i%
\mathbf{k}\cdot \mathbf{z}+2i\pi nz^{D}/L},\;n=0,\pm 1,\pm 2,\ldots .
\label{eigfuncD}
\end{equation}%
with the notations $\mathbf{k}=(k_{1},\ldots ,k_{D-1})$, $k=|\mathbf{k}|$,
and%
\begin{equation}
k_{n}=\sqrt{k^{2}+(2\pi n/L)^{2}},\;\nu =\left[ D^{2}/4-D(D+1)\xi
-m^{2}\alpha ^{2}\right] ^{1/2}.  \label{knD}
\end{equation}%
In (\ref{eigfuncD}) $H_{\nu }(x)$ is the Hankel function. The coefficient $%
C_{\sigma }$ with $\sigma =(\mathbf{k},n)$ is found from the
orthonormalization condition
\begin{equation}
-i\int d^{D}x\sqrt{|g|}g^{00}\left[ \varphi _{\sigma }(x)\partial
_{0}\varphi _{\sigma ^{\prime }}^{\ast }(x)-\varphi _{\sigma ^{\prime
}}^{\ast }(x)\partial _{0}\varphi _{\sigma }(x)\right] =\delta _{\sigma
\sigma ^{\prime }},  \label{normcond}
\end{equation}%
where the integration goes over the spatial hypersurface $\eta =\mathrm{const%
}$, and $\delta _{\sigma \sigma ^{\prime }}$ is understood as the Kronecker
delta for discrete indices and as the Dirac delta-function for continuous
ones. This leads to the result%
\begin{equation}
C_{\sigma }^{2}=\frac{\alpha ^{1-D}e^{-i(\nu -\nu ^{\ast })\pi /2}}{%
2^{D+1}\pi ^{D-2}L}.  \label{normCD}
\end{equation}

Substituting the eigenfunctions (\ref{eigfuncD}) with the normalization
coefficient (\ref{normCD}) into the mode-sum formula for the Hadamard
function, one finds
\begin{eqnarray}
G^{(1)}(x,x^{\prime }) &=&\frac{\alpha ^{1-D}(\eta \eta ^{\prime })^{D/2}}{%
2^{D}\pi ^{D-2}L}\int d\mathbf{k}\,e^{i\mathbf{k}\cdot \Delta \mathbf{z}}%
\sideset{}{'}{\sum}_{n=0}^{\infty }\cos (2\pi n\Delta z^{D}/L)  \notag \\
&&\times \left[ H_{\nu }^{(2)}(k_{n}\eta )H_{\nu }^{(1)}(k_{n}\eta ^{\prime
})+H_{\nu }^{(2)}(k_{n}\eta ^{\prime })H_{\nu }^{(1)}(k_{n}\eta )\right] ,
\label{GxxD}
\end{eqnarray}%
where $\Delta z^{D}=z^{D}-z^{\prime D}$. For the further evaluation we apply
to the series over $n$ the Abel-Plana summation formula \cite{Most97,Saha00}%
\begin{equation}
\sideset{}{'}{\sum}_{n=0}^{\infty }f(n)=\int_{0}^{\infty
}dx\,f(x)+i\int_{0}^{\infty }dx\,\frac{f(ix)-f(-ix)}{e^{2\pi x}-1},
\label{Abel}
\end{equation}%
where the prime on the summation sign means that the term $n=0$ should be
halved. This enables us to present the Hadamard function in the decomposed
form
\begin{equation}
G^{(1)}(x,x^{\prime })=G_{0}^{(1)}(x,x^{\prime })+G_{c}^{(1)}(x,x^{\prime }),
\label{G1decomp}
\end{equation}%
with%
\begin{eqnarray}
G_{0}^{(1)}(x,x^{\prime }) &=&\frac{(\eta \eta ^{\prime })^{D/2}}{2^{D+2}\pi
^{D-1}\alpha ^{D-1}}\int d\mathbf{k}_{D}\,e^{i\mathbf{k}_{D}\cdot \Delta
\mathbf{z}_{D}}  \notag \\
&&\times \left[ H_{\nu }^{(2)}(k_{D}\eta )H_{\nu }^{(1)}(k_{D}\eta ^{\prime
})+H_{\nu }^{(2)}(k_{D}\eta ^{\prime })H_{\nu }^{(1)}(k_{D}\eta )\right] ,
\label{G0xxp}
\end{eqnarray}%
being the Hadamard function for the uncompactified dS spacetime. In formula (%
\ref{G1decomp}), the second term on the right is induced by the compactness
of the $z^{D}$ - direction and is given by the formula

\begin{eqnarray}
G_{c}^{(1)}(x,x^{\prime }) &=&\frac{2(\eta \eta ^{\prime })^{D/2}}{(2\pi
)^{D}\alpha ^{D-1}}\int d\mathbf{k}\,e^{i\mathbf{k}\cdot \Delta \mathbf{z}%
}\int_{0}^{\infty }dx\,\frac{x\cosh (\sqrt{x^{2}+k^{2}}\Delta z_{D})}{\sqrt{%
x^{2}+k^{2}}(e^{L\sqrt{x^{2}+k^{2}}}-1)}  \notag \\
&&\times \left\{ \left[ I_{-\nu }(x\eta ^{\prime })+I_{\nu }(x\eta ^{\prime
})\right] K_{\nu }(x\eta )+\left[ I_{-\nu }(x\eta )+I_{\nu }(x\eta )\right]
K_{\nu }(x\eta ^{\prime })\right\} .  \label{GxxD2}
\end{eqnarray}%
Note that in this formula the integration with respect to the angular part
of $\mathbf{k}$ can be done explicitly. Two-point function in the
uncompactified dS spacetime given by formula (\ref{G0xxp}) can be expressed
in terms of the hypergeometric function and is investigated in \cite%
{Cand75,Dowk76,Dowk76b,Bunc78} (see also \cite{Birr82}).

\section{Vacuum expectation values of the field square and the
energy-momentum tensor}

\label{sec:vevEmt}

\subsection{Field square}

The VEV of the field square is obtained from the two-point function $%
G^{(1)}(x,x^{\prime })$ taking the coincidence limit of the arguments. In
this limit the Hadamard function diverges and some renormalization procedure
is necessary. The important point here is that the divergences are contained
in the part corresponding to the uncompactified dS spacetime and the
topological part is finite. As we have already extracted the first part, the
renormalization procedure is reduced to the renormalization of the
uncompactified dS part which is already done in literature. As a result the
renormalized VEV of the field square is presented in the form
\begin{equation}
\langle \varphi ^{2}\rangle _{\mathrm{ren}}=\langle \varphi ^{2}\rangle _{0,%
\mathrm{ren}}+\langle \varphi ^{2}\rangle _{c},  \label{phi2decomp}
\end{equation}%
where in the case $D=3$ the renormalized VEV for the uncompactified dS space
is given by the formula \cite{Cand75,Dowk76b,Bunc78}
\begin{eqnarray}
\langle \varphi ^{2}\rangle _{0,\mathrm{ren}} &=&\frac{1}{8\pi ^{2}\alpha
^{2}}\left\{ \left( m^{2}\alpha ^{2}/2+6\xi -1\right) \left[ \psi \left(
\frac{3}{2}+\nu \right) +\psi \left( \frac{3}{2}-\nu \right) -\ln \left(
m^{2}\alpha ^{2}\right) \right] \right.  \notag \\
&&\left. -\frac{\left( 6\xi -1\right) ^{2}}{m^{2}\alpha ^{2}}+\frac{1}{%
30m^{2}\alpha ^{2}}-6\xi +\frac{2}{3}\right\} ,  \label{phi20ren}
\end{eqnarray}%
where $\psi (x)$ is the logarithmic derivative of the gamma-function. Due to
the maximal symmetry of the dS spacetime this VEV\ does not depend on the
spacetime point.

The second term on the right of Eq. (\ref{phi2decomp}) is the part due to
the compactness of the $z^{D}$ - direction. This part is directly obtained
from (\ref{GxxD2}) taking the coincidence limit of the arguments:
\begin{eqnarray}
\langle \varphi ^{2}\rangle _{c} &=&\frac{\alpha ^{1-D}\eta ^{D}}{2^{D-2}\pi
^{\frac{D+1}{2}}\Gamma \left( \frac{D-1}{2}\right) }\int_{0}^{\infty
}dk\,k^{D-2}  \notag \\
&&\times \int_{0}^{\infty }dx\,\frac{x\left[ I_{-\nu }(x\eta )+I_{\nu
}(x\eta )\right] }{\sqrt{x^{2}+k^{2}}(e^{L\sqrt{x^{2}+k^{2}}}-1)}K_{\nu
}(x\eta ).  \label{phi2D}
\end{eqnarray}%
Here $I_{\nu }(x)$ and $K_{\nu }(x)$ are the modified Bessel functions.
Introducing a new integration variable $y=\sqrt{x^{2}+k^{2}}$ and expanding $%
(e^{Ly}-1)^{-1}$, the integral over $y$ is easily evaluated and one finds%
\begin{equation}
\langle \varphi ^{2}\rangle _{c}=\frac{\alpha ^{1-D}}{2^{D/2-1}\pi ^{D/2+1}}%
\sum_{n=1}^{\infty }\int_{0}^{\infty }dx\,x^{D-1}\left[ I_{-\nu }(x)+I_{\nu
}(x)\right] K_{\nu }(x)\frac{K_{D/2-1}(nLx/\eta )}{(nLx/\eta )^{D/2-1}}.
\label{phi2D1}
\end{equation}%
The integral in this formula can be expressed in terms of the hypergeometric
function $_{4}F_{3}$. By taking into account the relation between the
conformal and synchronous time coordinates, we see that the VEV of the field
square is a function of the combination $L/\eta =Le^{t/\alpha }/\alpha $.
For a conformally coupled massless scalar field one has $\nu =1/2$ and $%
\left[ I_{-\nu }(x)+I_{\nu }(x)\right] K_{\nu }(x)=1/x$. The corresponding
integral in (\ref{phi2D1}) is explicitly evaluated and we find%
\begin{equation}
\langle \varphi ^{2}\rangle _{c}=\frac{\zeta _{\mathrm{R}}(D-1)}{2\pi
^{(D+1)/2}}\left( \frac{\eta }{\alpha L}\right) ^{D-1}\Gamma \left( \frac{D-1%
}{2}\right) ,\;\xi =\xi _{D},\;m=0,  \label{phi2Conf}
\end{equation}%
where $\zeta _{\mathrm{R}}(x)$ is the Riemann zeta function. We could obtain
this result directly from the corresponding result in the Minkowski
spacetime compactified along the direction $z^{D}$, by using the fact that
two problems are conformally related: $\langle \varphi ^{2}\rangle
_{c}=(\eta /\alpha )^{D-1}\langle \varphi ^{2}\rangle _{c}^{\mathrm{(M)}}$.

Let us consider the behavior of the topological part $\langle \varphi
^{2}\rangle _{c}$ in the VEV of the field square in the asymptotic regions
of the ratio $L/\eta $. For small values of this ratio, $L/\eta \ll 1$, we
introduce a new integration variable $y=Lx/\eta $. By taking into account
that for large values $x$ one has $\left[ I_{-\nu }(x)+I_{\nu }(x)\right]
K_{\nu }(x)\approx 1/x$, we find that to the leading order $\langle \varphi
^{2}\rangle _{c}$ coincides with the corresponding result for a conformally
coupled massless field, given by (\ref{phi2Conf}):%
\begin{equation}
\langle \varphi ^{2}\rangle _{c}\approx \frac{\zeta _{\mathrm{R}}(D-1)}{2\pi
^{(D+1)/2}}\left( \frac{\eta }{\alpha L}\right) ^{D-1}\Gamma \left( \frac{D-1%
}{2}\right) ,\;L/\eta \ll 1.  \label{phi2small}
\end{equation}%
In the opposite limit of large values for the ratio $L/\eta $, again we
introduce a new integration variable $y=Lx/\eta $ and expand the integrand.
For real values $\nu $ to the leading order we find%
\begin{equation}
\langle \varphi ^{2}\rangle _{c}\approx \frac{\left( \eta /L\right) ^{D-2\nu
}\Gamma (\nu )}{2\pi ^{D/2+1}\alpha ^{D-1}}\zeta _{\mathrm{R}}(D-2\nu
)\Gamma \left( D/2-\nu \right) ,\;\eta /L\ll 1.  \label{phi2large}
\end{equation}%
For pure imaginary values $\nu $ by a similar way we can see that%
\begin{equation}
\langle \varphi ^{2}\rangle _{c}\approx \frac{\alpha ^{1-D}(\eta /L)^{D}}{%
\pi ^{D/2}\sinh (|\nu |\pi )}{\mathrm{Im}}\left[ \frac{\Gamma \left(
D/2-i|\nu |\right) }{\Gamma (1-i|\nu |)}\zeta _{\mathrm{R}}(D-2i|\nu
|)(L/\eta )^{2i|\nu |}\right] ,\;\eta /L\ll 1.  \label{phi2largeb}
\end{equation}%
By taking into account relation (\ref{eta}) between synchronous and
conformal time coordinates, the corresponding formula can also be written in
the form%
\begin{equation}
\langle \varphi ^{2}\rangle _{c}\approx \frac{\alpha a_{D}(|\nu
|)e^{-Dt/\alpha }}{\pi ^{D/2}\sinh (|\nu |\pi )L^{D}}\sin \left[ 2|\nu
|t/\alpha +2|\nu |\ln (L/\alpha )+\phi _{0}\right] ,\;t\gg \alpha .
\label{phi2largeb1}
\end{equation}%
where $a_{D}(|\nu |)$ and $\phi _{0}$ are defined by the relation%
\begin{equation}
\frac{\Gamma \left( D/2-i|\nu |\right) }{\Gamma (1-i|\nu |)}\zeta _{\mathrm{R%
}}(D-2i|\nu |)=a_{D}(|\nu |)e^{i\phi _{0}}.  \label{defaD}
\end{equation}%
As we see, unlike to the case of real $\nu $, here the damping of the VEV
has an oscillatory nature.

\begin{figure}[tbph]
\begin{center}
\begin{tabular}{cc}
\epsfig{figure=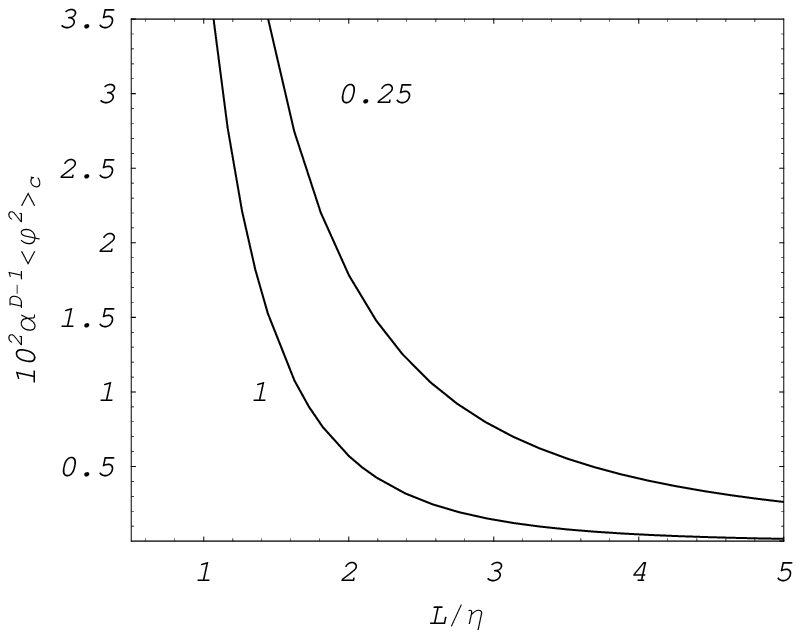,width=7.cm,height=6cm} & \quad %
\epsfig{figure=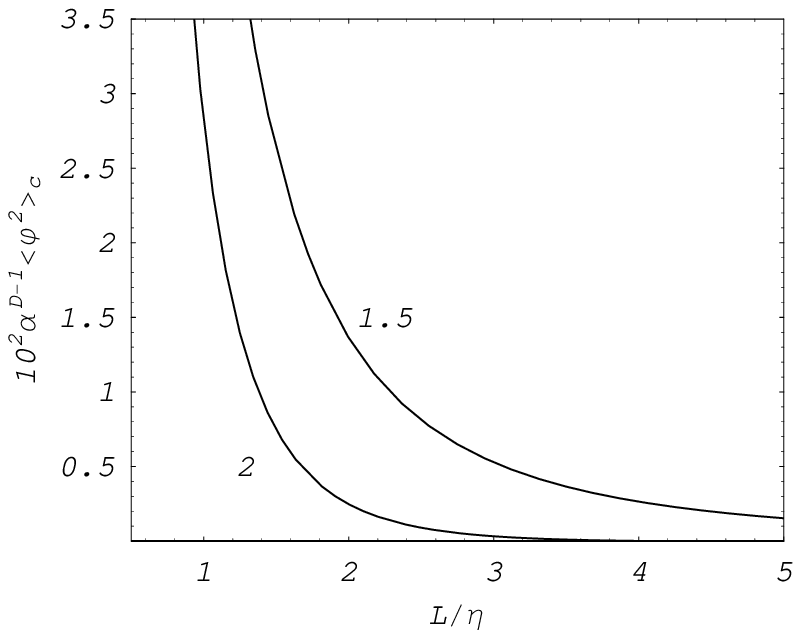,width=7.cm,height=6cm}%
\end{tabular}%
\end{center}
\caption{The topological part in the VEV of the field square for $D=3$
conformally (left panel) and minimally (right panel) coupled fields, $%
\protect\alpha ^{D-1}\langle \protect\varphi ^{2}\rangle _{c}$, as a
function of the ratio $L/\protect\eta $. The numbers near the curves
correspond to the values of the parameter $m\protect\alpha $. }
\label{fig1}
\end{figure}

In figure \ref{fig1} we have plotted $\langle \varphi ^{2}\rangle
_{c}$ as a function of the ratio $L/\eta $ for
conformally (left panel) and minimally (right panel) coupled fields in $D=3$%
. The numbers near the curves correspond to the value of the parameter $%
m\alpha $. Note that in the case $m\alpha =1$ for a conformally coupled
scalar the parameter $\nu $ is pure imaginary and the corresponding
oscillatory behavior is seen in the region $L/\eta >7$ (the first zero is at
$L/\eta \approx 7.28$). Similarly, for a minimally coupled scalar field the
parameter $\nu $ is pure imaginary in the case $m\alpha =2$ and the first
zero of $\langle \varphi ^{2}\rangle _{c}$ is at $L/\eta \approx 4.12$.

\subsection{Energy-momentum tensor}

\begin{figure}[tbph]
\begin{center}
\epsfig{figure=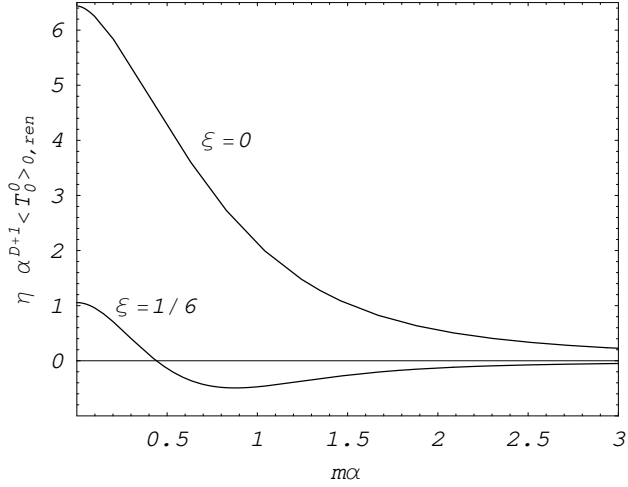,width=8.5cm,height=6.5cm}
\end{center}
\caption{Renormalized vacuum energy density in uncompactified dS spacetime, $%
\protect\eta \protect\alpha ^{D+1}\langle T_{0}^{0}\rangle _{\mathrm{ren}}$
as a function of $m\protect\alpha $ for minimally and conformally coupled
scalar fields in $D=3$. The scaling coefficient $\protect\eta =10^{3}(10^{4})
$ for minimally (conformally) coupled scalar fields.}
\label{fig2}
\end{figure}

Now we turn to the investigation of the VEV for the energy-momentum tensor.
Having the Hadamard function we can evaluate this VEV by making use of the
formula%
\begin{equation}
\langle 0|T_{ik}|0\rangle =\frac{1}{2}\lim_{x^{\prime }\rightarrow
x}\partial _{i}\partial _{k}^{\prime }G^{(1)}(x,x^{\prime })+\left[ \left(
\xi -\frac{1}{4}\right) g_{ik}\nabla _{l}\nabla ^{l}-\xi \nabla _{i}\nabla
_{k}-\xi R_{ik}\right] \langle 0|\varphi ^{2}|0\rangle ,  \label{emtvev1}
\end{equation}%
where $R_{ik}=Dg_{ik}/\alpha ^{2}$ is the Ricci tensor for the dS spacetime.
Similar to the case of the field square, the renormalized VEV of the
energy-momentum tensor is presented as the sum%
\begin{equation}
\langle T_{l}^{k}\rangle _{\mathrm{ren}}=\langle T_{l}^{k}\rangle _{0,%
\mathrm{ren}}+\langle T_{l}^{k}\rangle _{c},  \label{TikDecomp}
\end{equation}%
where $\langle T_{l}^{k}\rangle _{0,\mathrm{ren}}$ is the part corresponding
to the uncompactified dS spacetime and $\langle T_{l}^{k}\rangle _{c}$ is
induced by the compactness along the $z^{D}$ - direction. For $D=3$ the
standard dS part is given by the formula \cite{Cand75,Dowk76b,Bunc78} (see
also \cite{Birr82})
\begin{eqnarray}
\langle T_{l}^{k}\rangle _{0,\mathrm{ren}} &=&\frac{\delta _{l}^{k}}{32\pi
^{2}\alpha ^{4}}\left\{ m^{2}\alpha ^{2}\left( m^{2}\alpha ^{2}/2+6\xi
-1\right) \left[ \psi \left( 3/2+\nu \right) +\psi \left( 3/2-\nu \right)
-\ln \left( m^{2}\alpha ^{2}\right) \right] \right.   \notag \\
&&\left. -\left( 6\xi -1\right) ^{2}+1/30+(2/3-6\xi )m^{2}\alpha
^{2}\right\} ,  \label{Tik0ren}
\end{eqnarray}%
By using the asymptotic expansion of the function $\psi (x)$ for large
values of the argument it can be seen that for large values of the parameter
$m\alpha $ from (\ref{Tik0ren}) one has%
\begin{equation}
\langle T_{l}^{k}\rangle _{0,\mathrm{ren}}\approx \frac{\delta _{l}^{k}}{%
32\pi ^{2}m^{2}\alpha ^{6}}\left( \frac{7}{12}-\frac{58\xi }{5}+72\xi
^{2}-144\xi ^{3}\right) ,\;m\alpha \gg 1.  \label{Tik0renAs}
\end{equation}%
For a conformally coupled scalar field the coefficient in braces is equal $%
-1/60$. The energy-momentum tensor (\ref{Tik0ren}) is a gravitational source
of the cosmological constant type. Due to the problem symmetry this will be
the case for general values $D$. As a result, in combination with the
initial cosmological constant $\Lambda $ given by (\ref{Ricci}), the
one-loop effects lead to the effective cosmological constant (for a recent
discussion of the one-loop topological contributions to the cosmological
constant see \cite{Eliz06,Saha06})%
\begin{equation}
\Lambda _{\mathrm{eff}}=D(D-1)/2\alpha ^{2}+8\pi G\langle T_{0}^{0}\rangle
_{0,\mathrm{ren}},  \label{LambEff}
\end{equation}%
where $G$ is the Newton gravitational constant. In figure \ref{fig2} we have
plotted the renormalized vacuum energy density in the uncompactified dS
spacetime as a function of the parameter $m\alpha $ for $D=3$ minimally and
conformally coupled scalar fields. As it is seen from the plots, the
one-loop correction to the cosmological constant for a minimally coupled
scalar field is always positive, whereas for a conformally coupled scalar it
can be also negative. Below we will see that the energy density induced by
the compactness of a spatial dimension is not of cosmological constant type
and corresponds to an additional source in the cosmological equations which
is of barotropic type at late stages of the cosmological evolution.

The topological part in the VEV of the energy-momentum tensor is obtained
substituting the function (\ref{GxxD2}) into formula (\ref{emtvev1}). After
lengthy but straightforward calculations this leads to the result
\begin{eqnarray}
\langle T_{l}^{k}\rangle _{c} &=&\frac{2\delta _{l}^{k}}{\left( 2\pi \right)
^{D/2+1}\alpha ^{D+1}}\sum_{n=1}^{\infty }\int_{0}^{\infty }dx\,x^{D-1}\frac{%
K_{D/2-1}(nLx/\eta )}{(nLx/\eta )^{D/2-1}}F^{(l)}(x)  \notag \\
&&+\frac{4F_{0}^{(l)}}{\left( 2\pi \right) ^{D/2+1}\alpha ^{D+1}}%
\sum_{n=1}^{\infty }\int_{0}^{\infty }dx\,x^{D+1}\left[ I_{-\nu }(x)+I_{\nu
}(x)\right] K_{\nu }(x)\frac{K_{D/2}(nLx/\eta )}{(nLx/\eta )^{D/2}}.
\label{Tik1}
\end{eqnarray}%
where we have used the notations%
\begin{equation}
F_{0}^{(0)}=0,\;F_{0}^{(\beta )}=-1,\;F_{0}^{(D)}=(D-1),\;\beta =1,\ldots
,D-1,  \label{Fi}
\end{equation}%
and%
\begin{eqnarray}
F^{(0)}(x) &=&\left[ I_{-\nu }(x)+I_{\nu }(x)\right] K_{\nu }(x)\left( \nu
^{2}+2m^{2}\alpha ^{2}-x^{2}\right) +x^{2}\left[ I_{-\nu }^{\prime
}(x)+I_{\nu }^{\prime }(x)\right] K_{\nu }^{\prime }(x)  \notag \\
&&+D(1/2-2\xi )x\left[ (I_{-\nu }(x)+I_{\nu }(x))K_{\nu }(x)\right] ^{\prime
},  \label{F0} \\
F^{(\beta )}(x) &=&\left( 4\xi -1\right) \left[ I_{-\nu }(x)+I_{\nu }(x)%
\right] K_{\nu }(x)\left( x^{2}+\nu ^{2}\right) +\left( 4\xi -1\right)
x^{2}K_{\nu }^{\prime }(x),  \notag \\
&&\times \left[ I_{-\nu }^{\prime }(x)+I_{\nu }^{\prime }(x)\right] +\left[
2(D+1)\xi -D/2\right] x\left( \left[ I_{-\nu }(x)+I_{\nu }(x)\right] K_{\nu
}(x)\right) ^{\prime },  \label{F1} \\
F^{(D)}(x) &=&2x^{2}K_{\nu }(x)\left[ I_{-\nu }(x)+I_{\nu }(x)\right]
+F^{(1)}(x).  \label{FD}
\end{eqnarray}%
An alternative expressions for the topological part of the VEV of the
energy-momentum tensor are obtained from (\ref{Tik1}) after the integration
by parts:%
\begin{equation}
\langle T_{l}^{k}\rangle _{c}=\frac{4\alpha ^{-1-D}\delta _{l}^{k}}{\left(
2\pi \right) ^{D/2+1}}\sum_{n=1}^{\infty }\int_{0}^{\infty }dx\,x^{D-1}\left[
I_{-\nu }(x)+I_{\nu }(x)\right] K_{\nu }(x)f^{(l)}(x,nLx/\eta ),  \label{Tkl}
\end{equation}%
with the notations%
\begin{eqnarray}
f^{(0)}(x,u) &=&\frac{K_{D/2+1}(u)}{4u^{D/2-3}}-\left( D\xi +\frac{D+2}{4}%
\right) \frac{K_{D/2}(u)}{u^{D/2-2}}  \notag \\
&&+\left( m^{2}\alpha ^{2}-x^{2}+D^{2}\xi \right) \frac{K_{D/2-1}(u)}{%
u^{D/2-1}},  \label{f0} \\
f^{(\beta )}(x,u) &=&(\xi -1/4)\frac{K_{D/2+1}(u)}{u^{D/2-3}}-x^{2}\frac{%
K_{D/2}(u)}{u^{D/2}}-\xi D\frac{K_{D/2-1}(u)}{u^{D/2-1}}  \notag \\
&&-\left[ (D+1)\xi -\frac{D+2}{4}\right] \frac{K_{D/2}(u)}{u^{D/2-2}}%
,\;\beta =1,\ldots ,D-1,  \label{fbeta} \\
f^{(D)}(x,u) &=&(D-1)x^{2}\frac{K_{D/2}(u)}{u^{D/2}}+\left( x^{2}-\xi
D\right) \frac{K_{D/2-1}(u)}{u^{D/2-1}}+(\xi -1/4)\frac{K_{D/2+1}(u)}{%
u^{D/2-3}}  \notag \\
&&-\left[ (D+1)\xi -\frac{D+2}{4}\right] \frac{K_{D/2}(u)}{u^{D/2-2}}.
\label{fD}
\end{eqnarray}%
Note that the equivalent form (\ref{Tkl}) for the VEV of the energy-momentum
tensor can also be directly obtained from formula (\ref{emtvev1})
introducing a new integration variable $y=Lx/\eta $ in (\ref{phi2D1}) before
applying the differential operator. As it is seen from the obtained
formulae, the topological parts in the VEVs are time-dependent and, hence,
the local dS symmetry is broken by them.

It can be checked that the topological terms in the VEVs satisfy the trace
relation
\begin{equation}
\langle T_{i}^{i}\rangle _{c}=D(\xi -\xi _{D})\nabla _{l}\nabla ^{l}\langle
\varphi ^{2}\rangle _{c}+m^{2}\langle \varphi ^{2}\rangle _{c}.
\label{tracerel}
\end{equation}%
In particular, from here it follows that the topological part is traceless
for a conformally coupled massless scalar field. The trace anomaly is
contained in the uncompactified dS part only. Of course, we could expect
this result, as the trace anomaly is determined by the local geometry and
the local geometry is not changed by the compactification. For a conformally
coupled massless scalar field, similar to the case of the field square we
find (no summation over $l$)
\begin{eqnarray}
\langle T_{l}^{l}\rangle _{c} &=&-\frac{\zeta _{\mathrm{R}}(D+1)}{\pi
^{(D+1)/2}}\left( \frac{\eta }{\alpha L}\right) ^{D+1}\Gamma \left( \frac{D+1%
}{2}\right) ,\;l=0,1,\ldots ,D-1,  \label{TiiConf} \\
\langle T_{D}^{D}\rangle _{c} &=&-D\langle T_{0}^{0}\rangle _{c},\;\xi =\xi
_{D},\;m=0.  \label{TDDConf}
\end{eqnarray}%
Again, this result can be directly obtained by using the conformal relation
between the problem under consideration and the corresponding problem in the
Minkowski spacetime.

Now we turn to the investigation of the topological part in the asymptotic
regions of the parameters. For small values of the ratio $L/\eta $ we can
see that to the leading order $\langle T_{l}^{k}\rangle _{c}$ coincides with
the corresponding result for a conformally coupled massless field given by (%
\ref{TiiConf}), (\ref{TDDConf}):%
\begin{equation}
\langle T_{l}^{k}\rangle _{c}\approx \langle T_{l}^{k}\rangle _{c}(\xi =\xi
_{D},m=0),\;L/\eta \ll 1.  \label{TiiSmall}
\end{equation}%
In terms of the synchronous time coordinate this limit corresponds to $%
Le^{t/\alpha }\ll \alpha $. Now we see that in this limit the topological
part dominates the effective cosmological constant given by (\ref{LambEff})
and the back-reaction of one-loop quantum effects is important. From
formulae (\ref{TiiConf}), (\ref{TDDConf}) it follows that $\langle
T_{0}^{0}\rangle _{c}<0$, $\langle T_{0}^{0}\rangle _{c}-\langle
T_{D}^{D}\rangle _{c}<0$, \ $\langle T_{0}^{0}\rangle
_{c}-\sum_{i=1}^{D}\langle T_{i}^{i}\rangle _{c}<0$, and this part violates
all (weak, null, strong, dominant) energy conditions.

For large values of the ratio $L/\eta $ and in the case of real $\nu $ the
part $\langle T_{l}^{k}\rangle _{c}$ has the leading behavior given by the
formula (no summation over $l$)%
\begin{equation}
\langle T_{l}^{l}\rangle _{c}\approx \frac{\alpha ^{-1-D}}{2\pi ^{D/2+1}}%
\left( \frac{\eta }{L}\right) ^{D-2\nu }\zeta _{\mathrm{R}}(D-2\nu )\Gamma
(\nu )\Gamma \left( D/2-\nu \right) f_{0}^{(l)}(\nu ),\;\eta /L\ll 1,
\label{TiiLarge}
\end{equation}%
with the notations%
\begin{eqnarray}
f_{0}^{(0)}(\nu ) &=&2\nu D\left( \xi -\frac{D-2\nu }{4D}\right)
+m^{2}\alpha ^{2},  \label{f00} \\
f_{0}^{(l)}(\nu ) &=&-2\nu (D+1-2\nu )\left[ \xi -\frac{D-2\nu }{4(D+1-2\nu )%
}\right] ,\;l=1,2,\ldots D.  \label{f0i}
\end{eqnarray}%
In this limit the topological part corresponds to an effective gravitational
source with barotropic equation of state. In the same limit and for
imaginary $\nu $, to the leading order we have the following asymptotic:%
\begin{equation}
\langle T_{l}^{l}\rangle _{c}\approx \frac{\alpha ^{-1-D}(\eta /L)^{D}}{\pi
^{D/2}\sinh (|\nu |\pi )}{\mathrm{Im}}\left[ \frac{\Gamma \left( D/2-i|\nu
|\right) }{\Gamma (1-i|\nu |)}\zeta _{\mathrm{R}}(D-2i|\nu |)(L/\eta
)^{2i|\nu |}f_{0}^{(l)}(i|\nu |)\right] ,\;\eta /L\ll 1.  \label{TiiLargeIm}
\end{equation}%
As we see in both cases the leading terms in the vacuum stresses are
isotropic. The latter relation can also be written in terms of the
synchronous time coordinate:%
\begin{equation}
\langle T_{l}^{l}\rangle _{c}\approx \frac{a_{D}^{(l)}(|\nu |)e^{-Dt/\alpha }%
}{\pi ^{D/2}\sinh (|\nu |\pi )\alpha L^{D}}\sin \left[ 2|\nu |t/\alpha
+2|\nu |\ln (L/\alpha )+\phi _{0}^{(l)}\right] ,\;t\gg \alpha ,
\label{TiiLargeIm1}
\end{equation}%
where $a_{D}^{(l)}$ and $\phi _{0}^{(l)}$ are defined by the formula%
\begin{equation}
\frac{\Gamma \left( D/2-i|\nu |\right) }{\Gamma (1-i|\nu |)}\zeta _{\mathrm{R%
}}(D-2i|\nu |)f_{0}^{(l)}(i|\nu |)=a_{D}^{(l)}(|\nu |)e^{i\phi _{0}^{(l)}}.
\label{aDi}
\end{equation}

\begin{figure}[tbph]
\begin{center}
\begin{tabular}{cc}
\epsfig{figure=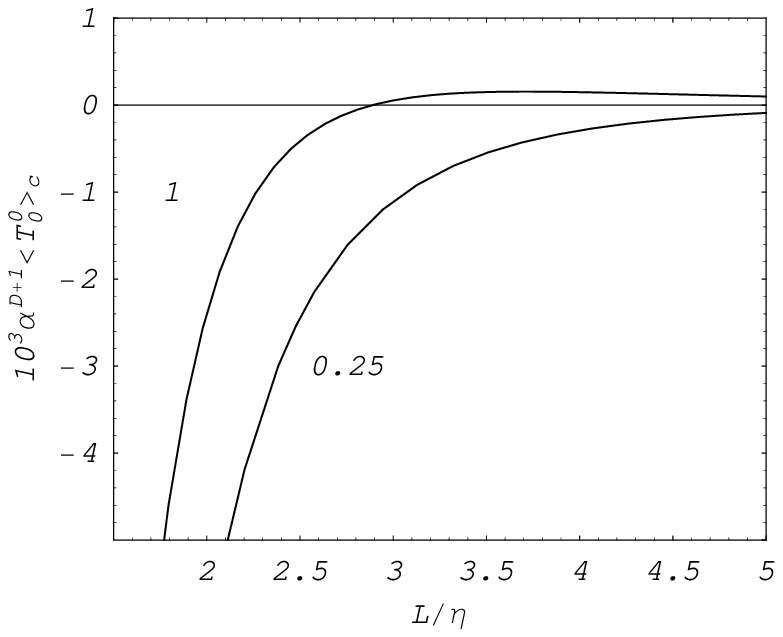,width=7.cm,height=6cm} & \quad %
\epsfig{figure=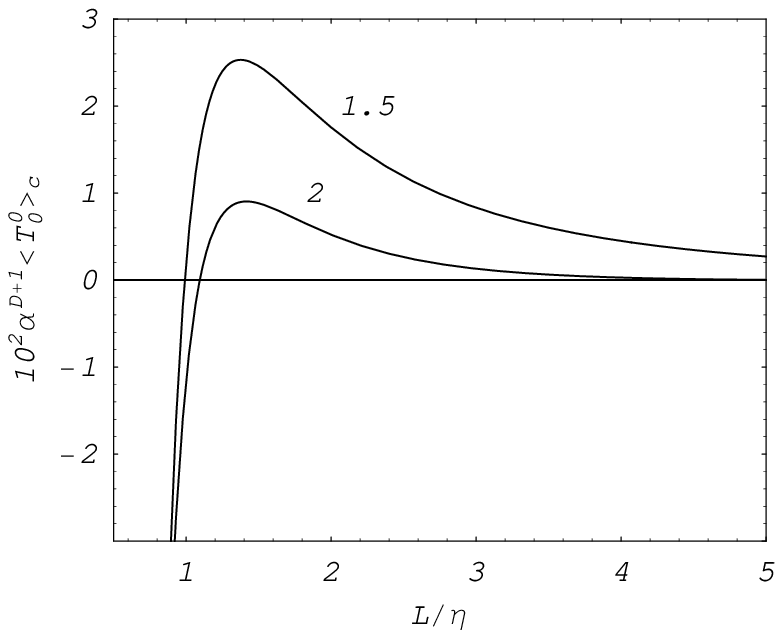,width=7.cm,height=6cm}%
\end{tabular}%
\end{center}
\caption{The part in the VEV of the energy density for conformally (left
panel) and minimally (right panel) coupled fields in $D=3$ induced by the
compactness of a spatial dimension, $\protect\alpha ^{D+1}\langle
T_{0}^{0}\rangle _{c}$, as a function of the ratio $L/\protect\eta $. The
numbers near the curves correspond to the values of the parameter $m\protect%
\alpha $.}
\label{fig3}
\end{figure}

In figure \ref{fig3} we have plotted $\langle T_{0}^{0}\rangle
_{c}$ as a function of the ratio $L/\eta $ for conformally (left
panel) and minimally (right panel) coupled fields in $D=3$. The
numbers near the curves correspond to the value of the parameter
$m\alpha $. The oscillatory behavior of the energy density in the
case $m\alpha =1$ for a conformally coupled scalar and in the case
$m\alpha =2$ for a minimally coupled scalar can be seen for larger
values of $L/\eta $.

\section{Conclusion}

\label{sec:Conc}

Compactified spatial dimensions appear in various physical models including
Kaluza-Klein type theories, supergravity, string theory and cosmology.
Motivated by the fact that dS spacetime plays an important role in all these
fields, in this paper we investigate the quantum vacuum effects in dS
spacetime induced by non-trivial topology of one of spatial dimensions. We
consider a scalar field with general curvature coupling parameter satisfying
the periodic boundary condition along the compactified dimension. Among the
most important characteristics of the vacuum are the VEVs of the field
square and the energy-momentum tensor. Though the corresponding operators
are local, due to the global nature of the vacuum these VEVs carry an
important information on the global structure of the background spacetime.

In order to derive formulae for the VEVs of the square of the field operator
and the energy-momentum tensor, we first construct the Hadamard function.
The application of the Abel-Plana summation formula enabled us to extract
from this function the part corresponding to the Hadamard function for the
uncompactified de Sitter spacetime. As the topological part is finite in the
coincidence limit, by this way the renormalization procedure is reduced to
that for the standard dS case. The latter was already realized in
literature. As a result the VEVs of the field square and the energy-momentum
tensor are decomposed into dS and topological parts. Due to the maximal
symmetry of dS spacetime the first one does not depend on the spacetime
point and is given by formula (\ref{phi20ren}) in the case of the field
square and by formula (\ref{Tik0ren}) for the energy-momentum tensor. This
is not the case for topological parts which are given by formulae (\ref%
{phi2D1}) and (\ref{Tkl}) for the field square and the energy-momentum
tensor respectively. In addition, the corresponding vacuum stresses along
uncompactified and compactified dimensions are anisotropic. As a result,
though the local geometry is not changed by the compactification, here we
deal with the topologically induced anisotropy in the properties of the
vacuum. Of course, this effect is well-known from the corresponding flat
spacetime examples. The topological terms take a simple form, given by
formulae (\ref{phi2Conf}), (\ref{TiiConf}), (\ref{TDDConf}), for special
case of a conformally coupled massless scalar field. These formulae can also
be obtained from the corresponding flat spacetime results by using the
conformal relation between the geometries.

The topological parts in the VEVs of the field square and the
energy-momentum tensor are functions of the ratio $L/\eta $. For general
values of the curvature coupling parameter the corresponding formulae are
simplified in the asymptotic regions of small and large values of this
parameter. In the first case the leading terms in the VEVs are the same as
those for a conformally coupled field and the topological parts behave like $%
\exp [-(D-1)t/\alpha ]$ for the field square and as $\exp [-(D+1)t/\alpha ]$
for the energy-momentum tensor. In this limit the topological part dominates
the effective cosmological constant and the back-reaction of one-loop
quantum effects is important. The corresponding energy-momentum tensor
violates the energy conditions and can essentially change the cosmological
dynamics. For large values of the ratio $L/\eta $ the behavior of the
topological parts is different for real and pure imaginary values of the
parameter $\nu $. In the first case these parts behave like $\exp [-(D-2\nu
)t/\alpha ]$, whereas in the second case the decay has an oscillatory nature
$e^{-Dt/\alpha }\sin \left( 2|\nu |t/\alpha +\phi _{1}\right) $. In this
limit the vacuum stresses are isotropic and the topological part corresponds
to an effective gravitational source with barotropic equation of state.

As it was argued in \cite{Lind04}, the models of a compact
universe with nontrivial topology may play an important role in
inflationary cosmology by providing proper initial conditions for
inflation. In particular, it was shown that in this class of
cosmological models the probability of inflation is not
exponentially suppressed. The quantum creation of the universe
having compact spatial topology is discussed in \cite{Zeld84} and
in \cite{Gonc85} within the framework of various supergravity
theories on non-simply connected backgrounds. Note that on
manifolds with non-trivial topologies or non-zero curvature, in
the evaluation of the vacuum energy no term-by-term cancellation
occurs between the bosonic and fermionic degrees of freedom (see,
for instance, \cite{Gonc85} for a flat spacetime with non-trivial
topology and \cite{Alle83} for the anti de Sitter bulk). In view
of this it will be interesting to generalize the results of the
present paper to other classes of the compactification, in
particular, for the toroidal compactification, including higher
spin fields. We will consider these topics in our future work.

\section*{Acknowledgments}

The work of AAS was supported in part by the Armenian Ministry of Education
and Science Grant No. 0124.

\end{document}